\documentclass{aa}

\usepackage{graphicx}
\usepackage{txfonts}
\usepackage{natbib}
\bibpunct{(}{)}{;}{a}{}{,}

\begin{document}

\title{Dynamics of magnetic flux tubes in close binary stars}
\subtitle{I.~Equilibrium and stability properties}
\titlerunning{Flux tubes in close binaries I.}

\author{V.~Holzwarth\inst{1,2} \and M.~Sch\"ussler\inst{1}}
\authorrunning{Holzwarth \& Sch\"ussler}

\institute{Max-Planck­Institut f\"ur Aeronomie, Max-Planck-Str.~2,
37191 Katlenburg-Lindau, Germany \\
\email{schuessler@linmpi.mpg.de}
\and 
School of Physics and Astronomy, University of St.~Andrews, North
Haugh, St.~Andrews KY16 9SS, UK \\
\email{vrh1@st-andrews.ac.uk}}

\date{Received gestern; accepted morgen}

\abstract{
Surface reconstructions of active close binary stars based on
photometric and spectroscopic observations reveal non-uniform starspot
distributions, which indicate the existence of preferred spot
longitudes (with respect to the companion star).
We consider the equilibrium and linear stability of toroidal magnetic
flux tubes in close binaries to examine whether tidal effects are
capable to initiate the formation of rising flux loops at preferred
longitudes near the bottom of the stellar convection zone.
The tidal force and the deviation of the stellar structure from
spherical symmetry are treated in lowest-order perturbation theory
assuming synchronised close binaries with orbital periods of a few
days.
The frequency, growth time, and spatial structure of linear eigenmodes
are determined by a stability analysis.
We find that, despite their small magnitude, tidal effects can lead to
a considerable longitudinal asymmetry in the formation probability of
flux loops, since the breaking of the axial symmetry due to the
presence of the companion star is reinforced by the sensitive
dependence of the stability properties on the stellar stratification
and by resonance effects.
The orientation of preferred longitudes of loop formation depends on
the equilibrium configuration and the wave number of the dominating
eigenmode.
The change of the growth times of unstable modes with respect to the
case of a single star is very small.

\keywords{ binaries: close -- stars: magnetic fields -- stars: spots --
stars: activity -- stars: imaging}
}

\maketitle

%
% Einleitung
%

\section{Introduction}
\label{intro}
Owing to their rapid rotation and deep convection zone, evolved
(sub-)giants and cool main sequence components in synchronised close
binaries like RS CVn and BY Dra systems typically show high levels of
magnetic activity \citep{1993A&AS..100..173S}, e.g., in the form of
large starspots and enhanced chromospheric and coronal emission in the
UV and X-ray spectral ranges.
Analyses of light curves with inversion techniques allow to derive the
characteristic properties of spots like their number, position, size,
and temperature \citep{1979ApJ...227..907E, 1986A&A...165..135R,
1987ApJ...319..827B}.
Utilising rotation-modulated, asymmetric spectral line profiles of
spectro- and spectropolarimetric observations, inversion techniques
like the Doppler Imaging \citep{1983PASP...95..565V,
1987ApJ...321..496V} and Zeeman-Doppler Imaging
\citep{1989A&A...225..456S, 1992A&A...265..682D, 1999MNRAS.302..437D}
reveal information about temperature and magnetic field inhomogeneities
on the stellar surface, which are interpreted as cool spots caused by
emerging magnetic fields.
Recent observations are complemented by photometric data which cover up
to several decades and thus enable the determination of cyclic
variations \citep{1995A&A...301...75R, 2000A&A...358..624R}.

Surface reconstructions of close, fast-rotating binaries show extended
starspots which cover large fractions of the stellar surface
\citep{1998ApJ...501L..73O, 2001A&A...376.1011L} and frequently occur,
in contrast to the case of the Sun, also at intermediate, high and even
polar latitudes.
Non-uniform spot distributions in longitude indicate the existence of
\emph{preferred longitudes} (PL), where spots occur more frequent or
last longer than at other longitudes \citep{1995ApJS...97..513H,
1996A&A...314..153J}.
Preferred longitudes are often found to be separated by about
$180\degr$, i.e., in opposite active quadrants of the hemisphere, but
their orientation with respect to the companion star is not unique.
In short-period systems with rotation periods $\lesssim 1\,\mathrm{d}$
active regions tend to prefer the quadrature longitudes
\citep{1989ApJ...345..991Z, 1990ApJ...363..647Z, 1990ApJ...354..352Z,
1994ApJ...421..303Z, 1994A&A...291..110O, 1998AJ....115.1145H}, i.e.,
the longitudes perpendicular to the line connecting both stellar
centres.
In systems with longer periods the situation is less clear: several
systems show PL at the substellar point, its antipode or other fixed
longitudes with respect to the companion star
\citep[e.g.,][]{1998A&A...332..541L, 2001A&A...376.1011L,
2002A&A...389..202O}, whereas other stars reveal a longitudinal
migration of the spotted regions with respect to the companion star,
which is usually ascribed to differential rotation
\citep{1994A&A...281..395S, 1995A&A...301...75R, 2000A&A...358..624R}.
Considering the dependence of PL on the orbital period,
\citet{1995AJ....109.2169H} find a transition from fixed PL for orbital
periods less that one day to migrating and eventually no PL at all for
more separated systems with periods exceeding about 10 days, leading to
the suggestion that tidal effects might influence the spot
distribution.
However, observations of close binaries with giant components show that
spot clusters at migrating PL may also occur in systems with longer
orbital periods \citep{1998A&A...336L..25B, 1999A&A...350..626B},
giving rise to the assumption that the formation and orientation of
clusters are not exclusively dependent on the binary separation alone.

In this and a subsequent paper we study the influence of tidal effects
on the dynamics and evolution of magnetic flux tubes in active binary
components, assuming -- in analogy to the case of the Sun -- that
starspots are formed by erupting flux tubes which originate from the
bottom of the stellar convection zone.
The present work delves into the equilibrium and linear stability
properties of flux tubes stored in the convective overshoot region
below the convection zone proper in order to examine whether the
influence of the companion star is able to trigger rising loops at
preferred longitudes which may penetrate to the convection zone above.
The decomposition of small perturbations of an equilibrium flux tube in
terms of eigenmodes and the determination of the corresponding
eigenfrequencies are accomplished by a linear stability analysis which
extends earlier treatments by \citet{1993GAFD...72..209,
1995GAFD...81..233} and \citet{2000AN....321..175H}.
In contrast to rotating single stars, in binary stars the existence of 
the companion star breaks the rotational symmetry. 
Consequently, the dynamics of flux tubes is not only subject to the
tidal force but is additionally affected by the deviation of the
stellar structure from spherical symmetry.
The linear stability analysis deals with the response of equilibrium
flux tubes to small perturbations inside the overshoot region, whereas
the growing displacements of unstable loops have to be followed by
non-linear numerical simulations.
The evolution of flux loops rising through the convection zone and the
resulting surface distribution upon their eruption is described in a
following paper.

In Sect.~\ref{modas} we briefly summarise the underlying (solar)
activity model based on erupting magnetic flux tubes and outline the
binary model.
Sect.~\ref{equi} describes the equilibrium properties and
Sect.~\ref{lisa} gives the results of the linear stability analysis of
toroidal flux tubes.
Sect.~\ref{disc} contains a discussion of the results and
Sect.~\ref{conc} gives our conclusions.

%
% Modellannahmen
%

\section{Model assumptions}
\label{modas}

\subsection{The `flux tube paradigm'}
\label{sopa}
It is widely accepted that sunspots and bipolar spot groups are caused
by erupting magnetic flux tubes which originate from the base of the
solar convection zone.
The magnetic field is believed to be amplified in the rotational shear
layer (tachocline) near the base of the convection zone and stored 
in the form of toroidal flux tubes in the convective overshoot region 
\citep{1992A&A...264..686M, 1994A&A...281L..69S}.
This stably stratified subadiabatic layer at the interface to the 
radiative core is generated by gas motions which penetrate from the 
convection zone above.
While Hale's polarity rules and the East-West-orientation of bipolar
sunspot groups suggest equilibrium flux rings parallel to the
equatorial plane, overshooting gas motions cause perturbations which
can trigger the onset of an undulatory (Parker-type) instability
\citep{1982A&A...106...58S, 1993GAFD...72..209, 1995GAFD...81..233}.
If a tube segment is lifted upward, a net mass downflow from the crest
increases the density contrast with respect to the environment.
Once a critical magnetic field strength is exceeded the resulting 
buoyancy surpasses the opposing magnetic curvature force, the loop 
continues to rise through the convection zone, and eventually erupts at 
the stellar surface \citep{1955ApJ...121..491P, 1986A&A...166..291M}.
Upon emergence it causes the various signatures of magnetic activity
like active regions, dark spot groups, bright faculae, chromospheric
activity, X-ray bright coronal loops, and flares.
The buoyancy-driven mechanism implies a crucial dependence on the 
stellar stratification, viz. the superadiabaticity, $\delta= \nabla -
\nabla_\mathrm{ad}$.
In the case of the Sun, magnetic flux tubes with field strengths of
about $10^5\,\mathrm{G}$ are required to obtain theoretical results
which are in accordance with observational constrains like flux
emergence latitudes, tilt angles, and asymmetries of bipolar spot
groups \citep{1992sto..work..385M, 1993A&A...272..621D,
1994ApJ...436..907F, 1994A&A...281L..69S, 1995ApJ...441..886C,
1998ApJ...502..481C}.
The quantitative elaboration of the applied flux tube model is also
consistent with the emergence of polar spots on rapidly rotating young
stars \citep{2000A&A...355.1087G} and with the conspicuous decline of
coronal X-ray emission in giant stars across the `coronal dividing
line' \citep{2001A&A...377..251H}.

We consider the magnetic flux in the convection zone in the form of
isolated magnetic flux tubes, which are embedded in a field-free
environment.
The stellar convection zone is treated as an ideal plasma with
vanishing viscosity and infinite conductivity, implying the concept of
frozen-in magnetic flux.
All calculations are carried out in the framework of the
`thin-flux-tube approximation' \citep{1978SoPh...56....5R,
1981A&A...102..129S}: the diameter of a circular flux tube is assumed
to be small with respect to the other relevant length scales of the
problem like the pressure scale height, radius of curvature, or the
wave length of a perturbation along the tube.
As a consequence of very short signal travel times it remains in
lateral (total) pressure equilibrium with its environment.  

%
% Doppelsternmodell
%

\subsection{Binary model}
Observations indicate that the orbital motion of binary systems with
periods $T\lesssim 10\,\mathrm{d}$ is effectively synchronised with
stellar rotation and typically exhibit very small eccentricity
\citep{1991A&A...248..485D, 1993A&AS..100..173S}.
Hence, we consider a close, detached, synchronised and circularised
binary system, for which the orbital and stellar rotation periods are
equal and the spin axes of the components are perpendicular to the
orbital plane.
The active star with mass $M_\star$ and the companion star with mass
$M_\mathrm{co}= q M_\star$ move on circular orbits with a period of a
few days.
According to Kepler's third law, they are separated by the constant 
distance 
\begin{equation}
\left( \frac{a}{R_{\sun}} \right)
\simeq
4.21\,
\left( 1 + q \right)^{1/3}
\left( \frac{M_\star}{M_{\sun}} \right)^{1/3}
\left( \frac{T}{\textrm{d}} \right)^{2/3}
\ .
\label{kepler}
\end{equation}
Our \emph{reference binary system} consists of two solar-type stars 
($M_\star= M_\mathrm{co}= 1\,M_{\sun}$, i.e. stellar mass ratio $q= 1$)
with a period of $T= 2\,\mathrm{d}$.
The active star is described by a perturbed single star (here, standard
solar) model, whereas the companion star is considered as an idealised 
point mass.
In the case of the Sun, the differential rotation profile affects the
stability properties only slightly
\citep[e.g.,][]{1993GAFD...72..209}.
Owing to tidal interactions and synchronisation, the differential
rotation in close binary stars is much weaker than in the solar case
\citep{1993PhDT.........3D, 1998ApJ...494..691R}, so that we assume
solid-body rotation.

\begin{figure}
\includegraphics[width=\hsize]{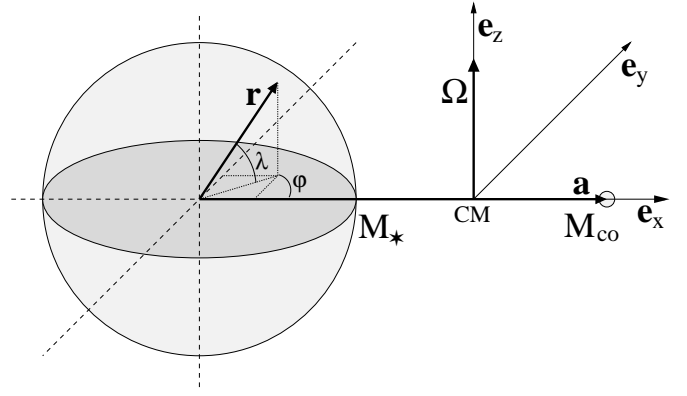}
\caption{
Geometry of the binary system.
The origin of the co-rotating frame of reference lies in the 
centre of mass, CM.
The stability analysis is carried out in spherical coordinates $(r,
\phi, \lambda)$ with respect to the centre of the active star,
$M_\star$.
}
\label{binko}
\end{figure}

For systems with orbital periods of a few days, tidal effects are 
sufficiently small to be treated in lowest-order perturbation theory.
Using spherical coordinates in a co-rotating frame of reference (radius
$r$, latitude $\lambda$, azimuth $\phi$, with the companion star in the
direction $\phi= 0$), the effective potential, $\Psi_\mathrm{eff}=
\Psi_\star + \Psi_\mathrm{tide}$, is the sum of the spherical
gravitational potential of the star, $\Psi_\star (r)$, and the
potential of the tidal perturbation\footnote{Equations (\ref{pottide}) 
and (\ref{gtide}) also contain the centrifugal potential and 
acceleration, respectively, although these contributions do not 
strictly belong to tidal interactions but occur in rotating single 
stars as well.},
\begin{eqnarray}
\Psi_\mathrm{tide} (r, \phi, \lambda)
& = &
-
\epsilon^3
\frac{G M_\star}{2 r}
\left[
 \left( 1 + \frac{5}{2} q \right)
 \cos^2 \lambda
 -
 q
\right. \nonumber \\ & & \left. {}
 +
 \frac{3}{2}
 q
 \cos^2 \lambda
 \cos 2\phi
\right]
+
\mathcal{O} \left( \epsilon^4 \right)
\ ,
\label{pottide}
\end{eqnarray}
where $\epsilon= r/a$ is the expansion parameter and $G$ the 
gravitational constant.
The potential $\Psi_\mathrm{tide}$ gives rise to the tidal acceleration
\begin{eqnarray}
\vec{g}_\mathrm{tide} (r, \phi, \lambda)
& = &
\epsilon^3
g_\star (r)
\left[
 \vec{e}_r
 +
 3 q
 \left( \vec{e}_r \cdot \vec{e}_a \right)
 \vec{e}_a
\right. \nonumber \\ & & \left. {}
 -
 \left( 1 + q \right)
 \left( \vec{e}_r \cdot \vec{e}_\Omega \right)
 \vec{e}_\Omega
\right]
+
\mathcal{O} \left( \epsilon^4 \right)
\ ,
\label{gtide}
\end{eqnarray}
where $\vec{e}_r,\vec{e}_\Omega$, and $\vec{e}_a$ are unit vectors
according to the directions indicated in Fig.~\ref{binko}.
Following Eq.~(\ref{gtide}), the component of the tidal acceleration
in azimuthal direction $\vec{e}_\phi$, for example, is
\begin{equation}
\left( \vec{g}_\mathrm{tide} \cdot \vec{e}_\phi \right) 
=
-
\frac{3}{2}
g_\star (r)
\epsilon^3
q
\cos \lambda
\sin 2 \phi
+
\mathcal{O} \left( \epsilon^4 \right)
\ .
\label{gtidephi}
\end{equation}

Our model includes the deviation of the stellar structure from
spherical symmetry arising from the tidal acceleration
$\vec{g}_\mathrm{tide}$.
For a not to strongly deformed star, we suppose that the stellar
structure is still described appropriately by functions of the
effective potential, $\Psi_\mathrm{eff}$, only, i.e.~$f(r, \phi,
\lambda)= f(\Psi_\mathrm{eff})$, where $f$ stands for any stellar
quantity (like gas pressure, density,$\ldots$).
This permits an analytical expression for the tidal deformation, namely
\begin{eqnarray}
f (r, \phi, \lambda)
& = &
f (\Psi_\star + \Psi_\mathrm{tide})
\approx
f (\Psi_\star)
+
\left.
 \frac{ \mathrm{d} f}{\mathrm{d} \Psi} 
\right|_r
\Psi_\mathrm{tide}
\nonumber \\ 
& = &
f_0
\left(
 1
 +
 \frac{r}{H_f} 
 \frac{\Psi_\mathrm{tide} r}{G M_\star}
\right)
\nonumber \\
& = &
f_0
\left(
 1
 +
 \bar{r}
 \frac{r}{H_f}
 +
 \hat{r}
 \frac{r}{H_f}
 \cos 2 \phi
\right)
\ ,
\label{fapprox}
\end{eqnarray}
with the dimensionless coefficients
\begin{eqnarray}
\bar{r}
& = &
\epsilon^3
\frac{1}{2} 
\left[ 
  \left(
    1 
    + 
    \frac{5}{2}
    q
  \right) 
  \cos^2 \lambda 
  - 
  q 
\right]
\ ,
\label{rbar}
\\
\hat{r}
& = &
\epsilon^3 
\frac{3}{4} 
q
\cos^2 \lambda 
\ ,
\label{rtilde}
\end{eqnarray}
and the scale height $H_f= - (\mathrm{d} \ln f/\mathrm{d} r)^{-1}$.
Equation (\ref{fapprox}) includes the rotational flattening and the 
tidal (ellipsoidal) deformation, with tidal bulges centred on the line
connecting the two stars.
The deviation of a quantity, $\Delta f= f(r,\phi,\lambda) - f_0 (r)$, 
from the single star stratification, $f_0$, depends on its local scale 
height, $H_f$.
Both values, $f_0$ and $H_f$, are taken from the spherical model at 
radius $r$.
The quality of the approximation can be estimated by considering the 
ratio between the geometrical deformation,
\begin{equation}
\frac{\delta r}{r}
\approx
\frac{\Psi_\mathrm{tide}}{\Psi_\star}
=
\bar{r}
+
\hat{r}
\cos 2 \phi
+
\mathcal{O} (\epsilon^4)
\ ,
\label{raddef}
\end{equation}
and the \emph{smallest} scale height $H_f$ of all quantities $f$.
Since in the overshoot region, the supposed storage location of 
magnetic flux, the scale height of the superadiabaticity, $H_\delta$, 
is considerably smaller than of any other quantity (see 
Tab.~\ref{charval}), the tidal influence on the stability properties
of flux tubes is mainly controlled by the variation of $\delta$ 
\citep{2000AN....321..175H}.

For our reference system ($T= 2\,\mathrm{d}, M_\star= M_{\sun}$ and $q=
1$), Eq.~(\ref{kepler}) yields a separation $a\sim 8\,R_{\sun}$ and 
$\epsilon^3\lesssim 10^{-3}$.
The tidal effects are of small magnitude only, but they imply a
significant \emph{qualitative} difference, since both
Eqs.~(\ref{pottide}) and (\ref{gtide}) exhibit a periodic azimuthal
dependence, which breaks the axial symmetry present in the single star
problem.

%
% Gleichgewicht
%

\section{Stationary equilibrium}
\label{equi}
\begin{figure}
\includegraphics[width=\hsize]{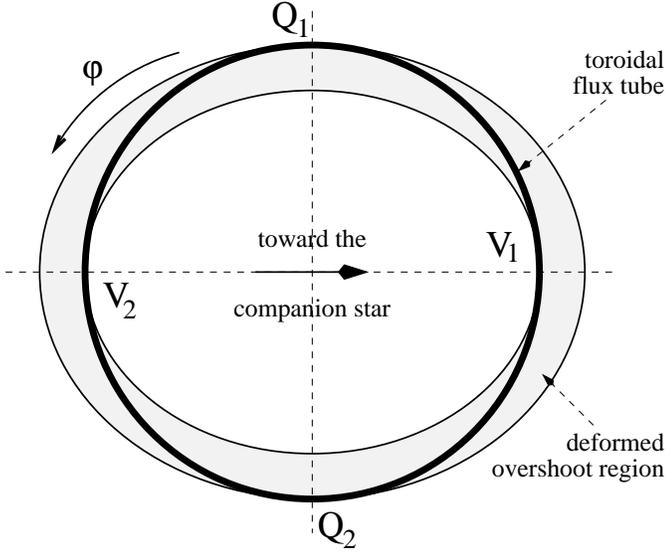}
\caption{
Sketch of a toroidal equilibrium flux tube embedded in the tidally 
deformed overshoot region.
The flux ring cuts through different equipotential surfaces, which 
implies a periodic variation of the environment along the tube: the 
values of the stratification at the points on the line of centres, 
$V_{1,2}$, correspond to deeper layers than those at the quadrature 
points, $Q_{1,2}$.
}
\label{defo.pic}
\end{figure}

We consider magnetic flux rings in mechanical stationary equilibrium, 
which are embedded in the overshoot region and parallel to the 
equatorial plane.
The tubes are non-buoyant and in lateral pressure equilibrium with the 
environment.
The inward directed magnetic curvature force is compensated by the 
outward directed Coriolis force, which arises from a prograde flow 
inside the tube with a velocity $v$ in excess of the stellar solid-body 
rotation.
The equilibrium tube forms a torus with constant radius of curvature, 
$R_0= r_0 \cos \lambda_0$, where $r_0$ and $\lambda_0$ are the radial 
and latitudinal equilibrium positions, respectively.
Since the flux ring is situated in a tidally deformed star 
(Fig.~\ref{defo.pic}), 
the internal density, $\rho$ ($= \rho_\mathrm{e}$, where 
$\rho_\mathrm{e}$ is the density of the environment), depends, 
according to the approximation in Eq.~(\ref{fapprox}), on the azimuth 
as
\begin{equation}
\frac{\rho (\phi)}{\rho_0} 
=
1
+
\bar{r}_0
\frac{r_0}{H_\mathrm{\rho0}}
+
\hat{r}_0
\frac{r_0}{H_\mathrm{\rho0}}
\cos 2 \phi
+
\mathcal{O} (\epsilon^4)
\ .
\label{rhovar}
\end{equation}
This entails a periodic variation of the magnetic field strength,
\begin{equation}
\frac{B (\phi)}{B_0}
=
\frac{
  1
  -
  M_{\alpha0}^2
}{
  \left( \frac{\rho (\phi)}{\rho_0} \right)
  - 
  M_{\alpha0}^2
}
\left( \frac{\rho (\phi)}{\rho_0} \right)
\ ,
\label{magvar}
\end{equation}
and internal flow velocity,
\begin{equation}
\frac{v (\phi)}{v_0}
=
\frac{
  1
  -
  M_{\alpha0}^2
}{
  \left( \frac{\rho (\phi)}{\rho_0} \right)
  - 
  M_{\alpha0}^2
}
\ ,
\label{velvar}
\end{equation}
where $M_{\alpha0}= v_0 / c_\mathrm{a0}$ is the \emph{Alfv\'enic Mach
number}, i.e., the flow velocity in units of the Alfv\'en velocity,
$c_\mathrm{a0}= B_0 / \sqrt{4\pi\rho_0}$.
Quantities with index `0' refer to a comparable axisymmetric
equilibrium flux tube in a slowly rotating single star, where the tidal
and centrifugal acceleration as well as the deformation of the stellar
structure are negligible \citep[e.g.,][]{1992A&A...264..686M}.
For flux tubes in mechanical equilibrium, these quantities obey the
relation 
\begin{equation}
v_0
=
R_0 \Omega
\left[
 \sqrt{
  1
  +
  \left( 
   \frac{c_\mathrm{a0}}{R_0\Omega} 
  \right)^2
 }
 -
 1
\right]
>
0
\ .
\label{vrel}
\end{equation}
In the following, we consider flux rings in the middle of the overshoot
region.
Table \ref{charval} gives values of the characteristic parameters of 
this reference configuration for a model of a $1\,M_{\sun}$ star.
\begin{table}
\centering
\caption{
Parameters at equilibrium depth $r_0$ of the reference configuration
}
\begin{tabular}{ll}
\hline
\hline
equilibrium depth & $r_0= 5.07\cdot10^{10}\,\mathrm{cm}= 
0.73\,R_{\sun}$ \\
gas pressure & $p_\mathrm{e0}= 4.31\cdot10^{13}\,\mathrm{dyn/cm^2}$ \\
pressure scale height & $H_{p0}= 5.52\cdot10^9\,\mathrm{cm}$ \\
density & $\rho_\mathrm{e0}= 0.15\,\mathrm{g/cm^3}$ \\
density scale height & $H_{\rho0}= 9.21\cdot10^9\,\mathrm{cm}$ \\
superadiabaticity & $\delta_0= -9.77\cdot10^{-7}$ \\
superad. scale height & $H_{\delta0}= 4.43\cdot10^8\,\mathrm{cm}$ \\
\hline
rotation & $T= 2\,\mathrm{d}$ \\
binary separation & $a= 8.41\,R_{\sun}$ \\
expansion parameter & $\epsilon^3= 6.53\cdot10^{-4}$ \\
deformation parameters & $\bar{r}= 8.16\cdot10^{-4}$ \\
(at equator, $\lambda= 0$) & $\hat{r}= 4.90\cdot10^{-4}$ \\
\hline
\end{tabular}
\label{charval}
\end{table}
%

%
% Linear stability analysis
%

\section{Linear stability analysis}
\label{lisa}

\subsection{Determination and interpretation of eigenmodes}
\label{thpr}
\begin{figure}
\includegraphics[width=\hsize]{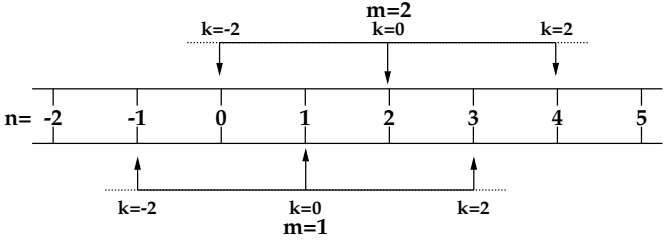}
\caption{
An eigenmode in the binary problem is represented by a spectrum of 
coupled wave modes with wave numbers $n= m\pm k, k=0,2,4,\ldots$
Due to the weak coupling ($\propto \epsilon^3$) the eigenmode can still
be characterised by the wave number $m$ of its dominating constituent.
The contributions of wave modes with $k\neq 0$ lead to non-axisymmetric
modifications, whereas wave modes with $|k|\ge 4$ (indicated by dots)
are negligible.
}
\label{kopplung.pic}
\end{figure}

Once all quantities of the external stratification are determined by
the equilibrium depth $r_0$, the stability properties of the flux ring
depend only on its latitude, $\lambda_0$, and magnetic field strength, 
$B_0$.
Lagrangian displacements are decomposed in eigenmodes,
\begin{equation}
\vec{\Xi} (\phi,t)
= 
\vec{\xi} (\phi) 
\exp \left( i \omega t \right)
\ ,
\label{xians}
\end{equation}
characterised by the eigenfrequency $\omega$ and the eigenfunction 
$\vec{\xi} (\phi)= (\xi_\mathrm{t}, \xi_\mathrm{n}, \xi_\mathrm{b})^T$ 
(in the co-moving trihedron with the local tangent, normal, and
binormal unit vectors, $[\vec{t}, \vec{n}, \vec{b}]$, respectively).
For unstable eigenmodes with increasing amplitudes $\tau= 1 / 
|\Im(\omega)|$ provides the characteristic growth time.
The displacement vector $\vec{\xi}$ is determined by the linearised 
equations of motion,
\begin{equation}
\vec{\xi}''
+
\left(
 \mathcal{M}_\phi
 +
 i
 \omega
 \mathcal{M}_{\phi t}
\right)
\vec{\xi}'
+
\left(
 \mathcal{M}_\xi
 +
 i
 \omega
 \mathcal{M}_t
 -
 \omega^2
 \mathcal{M}_{tt}
\right)
\vec{\xi}
=
0
\ ,
\label{hlinsys}
\end{equation}
where primes denote derivatives with respect to the azimuthal
coordinate, $\phi$ (see Appendix \ref{nume}).
Taking the tidal force, Eq.~(\ref{gtide}), and the azimuthal variation
of the equilibrium quantities, Eqs.~(\ref{rhovar}) -- (\ref{velvar}),
into account, the coefficient matrices depend periodically on $\phi$:
\begin{eqnarray}
\mathcal{M} (\phi)
& = &
\mathcal{M} (\phi + \pi)
\nonumber \\
& = &
\mathcal{M}_\mathrm{0} 
+
\epsilon^3
\mathcal{M}_\mathrm{c} 
\cos 2 \phi
+
\epsilon^3
\mathcal{M}_\mathrm{s} 
\sin 2 \phi
+
\mathcal{O} (\epsilon^4)
\ .
\label{matvar}
\end{eqnarray}
Closed flux tubes require $\vec{\xi} (\phi + 2\pi)\equiv \vec{\xi} 
(\phi)$ and $\vec{\xi}' (\phi + 2\pi)\equiv \vec{\xi}' (\phi)$, 
suggesting the ansatz
\begin{eqnarray}
\vec{\xi} (\phi)
& = &
\vec{\hat{\xi}} (\phi)
\exp \left( i m \phi \right)
\\
& = &
\left[
\sum_{k=-\infty}^{\infty}
\vec{\hat{\xi}}_k
\exp \left( i k \phi \right)
\right]
\exp \left( i m \phi \right)
\ .
\label{ampans}
\end{eqnarray}
In analogy to the single-star problem, a phase factor with the
azimuthal wave number $m$ has been separated.
Due to the $\phi$-dependent contributions in Eq.~(\ref{matvar}), the 
substitution of Eq.~(\ref{ampans}) into Eq.~(\ref{hlinsys}) yields 
the 3-term recursion formula
\begin{equation}
\mathcal{L}_{m+k} \vec{\hat{\xi}}_{k-2}
+
\mathcal{C}_{m+k} \vec{\hat{\xi}}_{k}
+
\mathcal{R}_{m+k} \vec{\hat{\xi}}_{k+2}
=
0
\quad , \qquad
\forall k
\ .
\label{recu}
\end{equation}
In contrast to the single-star problem, where an eigenmode corresponds
to a single azimuthal wave number $m$, i.e., $\vec{\hat{\xi}}_0=
\mathrm{const.}$ 
and $\vec{\hat{\xi}}_k= 0$ for $k\neq 0$, an eigenmode here consists of
a spectrum of coupled wave modes $n= m+k$ with $k= 0,\pm2,\pm4,\ldots$,
whose amplitudes $\vec{\hat{\xi}}_k$ are recursively related to each 
other through Eq.~(\ref{recu}).
This relation is illustrated in Fig.~\ref{kopplung.pic}.
Since the coupling is weak [$\propto \mathcal{O} (\epsilon^3)$], the
amplitudes decrease quite fast with increasing $|k|$, so that more
distant wave modes (here, with $|k|\ge 4$) do practically not
contribute to the eigenmode.
In most cases the amplitude $\vec{\hat{\xi}}_0$ is much larger than 
$\vec{\hat{\xi}}_{\pm2}$, so that the resulting eigenmode can still 
be characterised by the wave number $m$.

The linear eigenfunctions characterise the structure of small
displacements of flux tubes in the overshoot region \emph{prior} to
their rapid, non-linear rise through the convection zone.
Owing to the stellar stratification, the most important aspect of an 
eigenfunction is its component in the radial direction, 
$\hat{\xi}_r= (\vec{\hat{\xi}}\cdot\vec{e}_r)$: the larger the radial 
displacements, the larger are the changes in the environment of the 
corresponding tube segments, which become located in either more stable
or more unstable layers.
Assuming that larger radial displacements toward more unstable layers 
favour the penetration of a growing flux loop to the convection zone, 
we use the envelope, $|\hat{\xi}_r (\phi)|$, of the radial component of
the (normalised) eigenfunction as a measure for the onset of rapidly 
rising flux loops at azimuth $\phi$.
Since for eigenmodes with $\Im(\omega)< 0$ the magnitude of the radial 
envelope increases exponentially in time,
\begin{equation}
| \Xi_r (\phi,t)|
=
| \hat{\xi}_r (\phi) |
\exp \left[ - \Im(\omega) t \right]
\ ,
\label{pertmodu}
\end{equation}
an unstable flux tube in the overshoot region will eventually enter the 
superadiabatical region.
Its actual displacement depends, however, also on the azimuthal wave 
number $m$, the frequency $\Re(\omega)$, and the phase relation
\begin{equation}
\arg \Xi_r (\phi,t)
=
m \phi + \arg \hat{\xi}_r (\phi) + \Re(\omega) t
\ ,
\label{pertphase}
\end{equation}
corresponding to an azimuthal propagation of the tube's crest along the 
envelope, Eq.~(\ref{pertmodu}), with the phase velocity
\begin{equation}
v_\mathrm{p} (\phi)
=
-
\frac{\Re(\omega) R_0}{m}
\left(
 1 
 + 
 \frac{1}{m}\frac{\mathrm{d}}{\mathrm{d} \phi} \arg \hat{\xi}_r (\phi) 
\right)^{-1}
\ .
\label{vphase}
\end{equation}
The radial envelope thus represents a kind of `statistical weight', 
which is applicable either to a large ensemble of comparable flux tubes
with similar equilibrium properties and phase shifts evenly distributed
in the range $0\le \arg \Xi_r< 2\pi$, or to an individual flux tube 
with a growth time much longer than the wave frequency, $\tau\gg
2\pi/\Re(\omega)$.
If an eigenmode is dominated by the wave mode $m$ and only slightly 
modified by the adjacent wave modes $m\pm2$, i.e., $\hat{\xi}_{r,\pm2} 
\ll \hat{\xi}_{r,0}:= 1$, the radial envelope is approximately given by
\begin{eqnarray}
\left| \hat{\xi}_r (\phi) \right|
& = &
\left|
 \hat{\xi}_{r,0}
 +
 \hat{\xi}_{r,2}
 e^{ i 2 \phi }
 +
 \hat{\xi}_{r,-2}
 e^{ - i 2 \phi }
 +
 \ldots
\right|
\nonumber \\
& \simeq &
1
+
\left| \hat{\xi}_{r,2} \right|
\cos \left( 2 \phi + \alpha_2 \right)
+
\left| \hat{\xi}_{r,-2} \right|
\cos \left( 2 \phi - \alpha_{-2} \right)
\nonumber \\ & & {}
+ 
\mathcal{O} ( \hat{\xi}_{r,\pm2}^2 )
\ ,
\label{absform}
\end{eqnarray}
with $\alpha_{\pm2}= \arg \hat{\xi}_{r,\pm2}$.
The eigenfunctions thus inherit the properties of the underlying 
problem and exhibit a $\pi$-periodicity in azimuthal direction.
Assuming that the shape of $|\hat{\xi}_r (\phi)|$ represents a measure
for the onset of rising loops, Eq.~(\ref{absform}) implies the
existence of \emph{preferred longitudes} at the maxima of
$|\hat{\xi}_r|$.

% Eigenfrequenzen

\subsection{Growth times and eigenfrequencies}
\label{grti}
\begin{figure}
\includegraphics[width=\hsize]{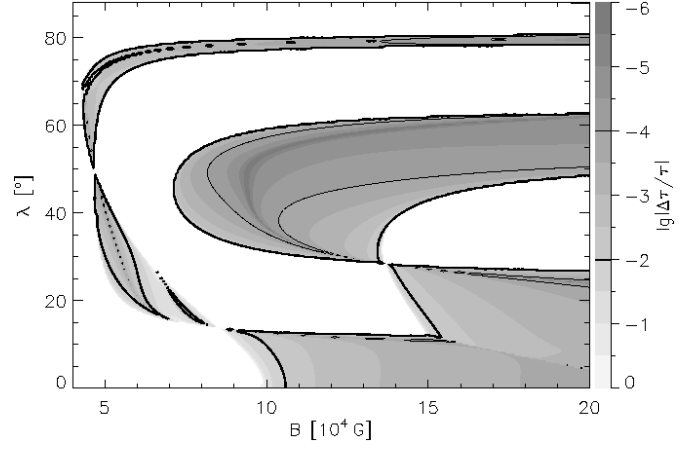}
\caption{
Relative deviation of the growth time, $(\tau_\mathrm{bin} - 
\tau_\mathrm{sin}) / \tau_\mathrm{sin}$, in parameter domains which 
have been unstable in the rotating single-star problem, i.e., 
$\tau_\mathrm{sin}\neq 0$.
}
\label{ef2d_tide_diff.pic}
\end{figure}

\begin{figure}
\includegraphics[width=\hsize]{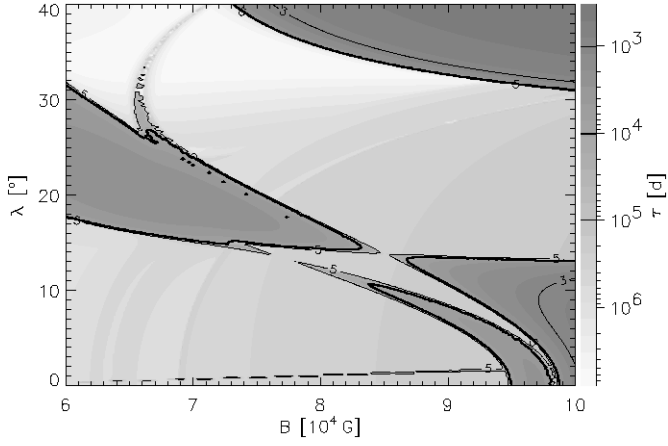}
\caption{
Growth times of the `instability background' due to tidal effects.
In addition to the Parker-type instabilities there are localised 
\emph{ridges} with $\tau\gtrsim 10^3\ldots10^5\,\mathrm{d}$ traversing 
extended \emph{plateaus} with $\tau> 10^5\,\mathrm{d}$.
Because of the limited resolution of the parameter grid in the figure 
the very narrow ridge at $\lambda\simeq 1\degr$ appears to be 
discontinuous.
}
\label{ef2d_box.pic}
\end{figure}

The influence of the companion star modifies the stability properties 
of flux tubes by breaking the axial symmetry.
The resulting \emph{quantitative} difference in growth times $\tau$ 
with respect to a rotating single star is very small; as illustrated in
Fig.~\ref{ef2d_tide_diff.pic}, the relative deviations in all cases 
are considerably smaller than $1\%$.
\emph{Qualitative} differences occur, however, for equilibrium 
configurations which had been stable in the single-star problem.
In previously stable parameter domains new types of instabilities with 
very long growth times lead to an `instability background' consisting 
of extended `plateaus' which are pervaded by narrow `ridges' (see 
Fig.~\ref{ef2d_box.pic}).

The `plateaus' are due to the tidal force and the azimuthal variation 
of the density along the equilibrium flux ring. 
Using the lateral (total) pressure equilibrium and the variation of the
magnetic field strength, Eq.~(\ref{magvar}), it can be shown that the
internal specific entropy, $S$, is also a function of $\phi$ and varies
approximately as 
\begin{eqnarray}
\frac{1}{c_p}
\frac{\partial S}{ \partial \phi}
& \sim &
\frac{1}{\beta}
\frac{\mathrm{d} \ln p_\mathrm{e}}{\mathrm{d} \phi}
\sim
\frac{\epsilon^3}{\beta}
\frac{r}{H_{p,{\rm e}}}
\ ,
\end{eqnarray}
where $\beta= 8\pi p/B^2$ is the ratio between the gas pressure and the
magnetic pressure, and $H_{p,\textrm{e}}$ is the local pressure scale
height.
Although we have $r/H_{p,\textrm{e}}\simeq 9$, the azimuthal variation 
is nevertheless small compared to the radial variation of the stellar 
stratification, since the relevant field strengths considered here
lead to values in the regime $\beta\sim 10^4\ldots10^6\gg 1$.
Due to its velocity excess, $v$, a gas element inside the tube would 
have to change its entropy to preserve the equilibrium conditions.
But since there is no mechanism available to accomplish this entropy 
change, the azimuthal variation leads to a loss of the mechanical 
equilibrium and to very slow monotonic displacements of the flux tube 
out of its toroidal configuration.
For fast-rotating stars, the internal flow velocity $v$ is small and
thus the growth times of plateau instabilities are typically several
$10^4$ days (up to $\tau\sim 10^7\,\mathrm{d}$).
The effects of the azimuthal density variation and tidal force along 
the tube are negligible because the stability properties are dominated 
by the azimuthal variation of the superadiabaticity,
\begin{equation}
\delta (\phi)
=
\delta (r_0)
\left(
 1
 +
 \frac{\delta r}{H_\delta}
\right)
\ ,
\label{deltavar}
\end{equation}
owing to its relative small scale height in the overshoot region (with 
$H_\delta/H_p\simeq 0.08$, see Tab.~\ref{charval}).
If all azimuthal variations other than that of $\delta$ are omitted in 
the stability analysis, the plateau instabilities vanish and only the 
ridge instabilities remain.

\begin{figure}
\includegraphics[width=\hsize]{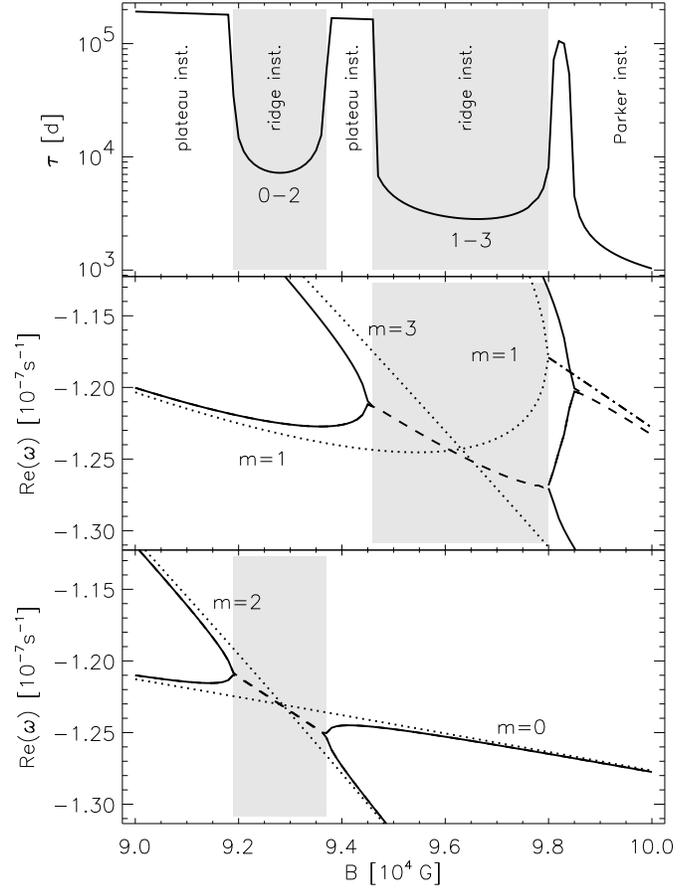}
\caption{
Growth times $\tau$ (\emph{top}) of Parker-type, ridge- and 
plateau-instabilities at $\lambda= 1.64\degr$ (see 
Fig.~\ref{ef2d_box.pic}).
The ridge instabilities (grey shading) are due to resonant 
interactions of adjacent wave modes with $m= 1$ and $3$ (\emph{middle})
and $m= 0$ and $2$ (\emph{bottom}), respectively, when the frequencies 
$\Re(\omega)$ are similar.
Dotted lines show the frequencies of the uncoupled modes in the 
corresponding single-star problem.
Dashed and dotted-dashed lines indicate unstable modes with 
$\Im(\omega)\neq 0$.
}
\label{ef2d_cut.pic}
\end{figure}

The `ridge' instabilities with typical growth times of several 
thousand days owe their existence to a resonant coupling of wave modes 
with different wave numbers.
As an example, the top panel of Fig.~\ref{ef2d_cut.pic} shows the
growth times of both Parker-type and background instabilities along a
cut at $\lambda= 1.64\degr$ in Fig.~\ref{ef2d_box.pic}.
The panels below show the frequencies $\Re(\omega)$ of the wave modes 
involved.
In the single-star problem (dotted lines, dashed-dotted if unstable)
the wave modes corresponding to different values of the azimuthal wave
number, $m$, are independent.
But the eigenmodes in the binary problem consist of a spectrum of 
\emph{coupled} wave modes (see Fig.~\ref{kopplung.pic}) with wave 
numbers $n=m, m\pm2, m\pm4,\ldots$: modes of wave number $m$ with
frequency $\Re(\omega)$ stimulate excitations of the wave modes 
$m\pm2$, which, supposed $\Re(\omega)$ is close to the eigenfrequency 
of the adjacent wave mode, results in a resonant amplification and 
eventually in the onset of an instability (solid lines in 
Fig.~\ref{ef2d_cut.pic}, dashed if unstable).

In the following, we dismiss instabilities with very long growth times 
$\tau> 10^4\,\mathrm{d}\;(\approx 27\,\mathrm{years})$, since they are 
not expected to play any role for the eruption of magnetic flux tubes 
in active stars.

% Eigenfunktionen

\subsection{Eigenfunctions}
\label{efct}
\begin{figure}
\includegraphics[width=\hsize]{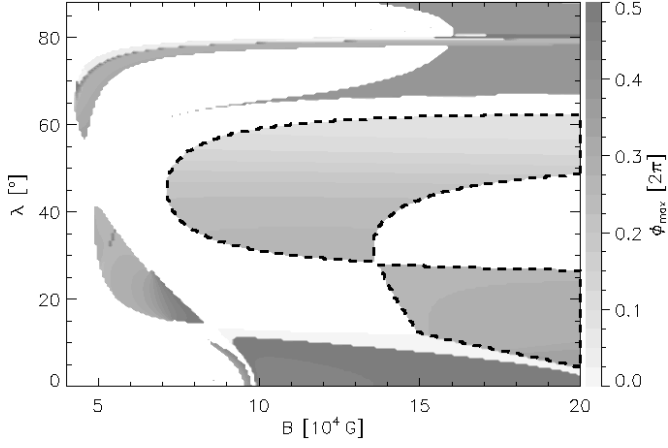}
\caption{
Azimuth $\phi_\mathrm{max}$ of the radial envelope maxima.
Due to the azimuthal $\pi$-periodicity only the maximum in the interval
$0\le \phi_\mathrm{max}< \pi$ is shown, i.e., between the points $V_1$ 
and $V_2$ on the line of centres implying the second maximum at 
$\phi_\mathrm{max} + \pi$.
The dashed framed region is the instability domain of eigenmodes with
a dominating wave number $m= 2$.
}
\label{ev2d_phase.pic}
\end{figure}

The eigenfunctions, $\vec{\hat{\xi}}$, inherit the property of the
underlying problem and exhibit a $\pi$-periodicity in azimuthal
direction, implying a $\phi$-dependent probability that a loop
penetrates to the superadiabatic part of the convection zone and rises
rapidly toward the surface.
The preferred longitudes, $\phi_\mathrm{max}$ (and $\phi_\mathrm{max} +
\pi$), of loop penetration are determined by the maxima of the radial
envelope, $|\hat{\xi}_r|_\mathrm{max}= |\hat{\xi}_r
(\phi_\mathrm{max})|$.
They are shown in Fig.~\ref{ev2d_phase.pic} as a function of the 
equilibrium latitude $\lambda_0$ and field strength $B_0$.
Background instabilities with growth times exceeding $10^4\,\mathrm{d}$
have been neglected, leaving behind only a fraction of a ridge at low 
latitudes and a plateau at $\lambda\gtrsim 60\degr$ with $\tau\gtrsim 
3\cdot10^3\,\mathrm{d}$.
This high-latitude feature is caused by the slow loss of equilibrium
due to the internal flow described above, which exhibits only small 
displacements in radial direction (here, $|\hat{\xi}_r| / 
|\hat{\xi}_\phi|\sim \mathcal{O}(10^{-2})$).
In contrast to the case of stable eigenfunctions, for which the 
envelope maxima are located exactly at the `symmetry points', $Q_{1,2}$
or $V_{1,2}$, the orientation of $|\hat{\xi}_r|_\mathrm{max}$ of
unstable eigenfunctions show a phase shift with respect to these
points.
For instabilities at equatorial and low latitudes dominated by the wave
number $m= 1$, the maxima are located in the vicinity of $V_{1,2}$, 
i.e., around the line connecting both stellar centres, whereas for 
modes with $m= 2$ they are distributed in a broad interval around the 
quadrature points $Q_{1,2}$; preferred longitudes of loop penetration
thus depend on the equilibrium parameters $(B_0,\lambda_0)$ as well as 
on the dominating wave mode.

\begin{figure}
\includegraphics[width=\hsize]{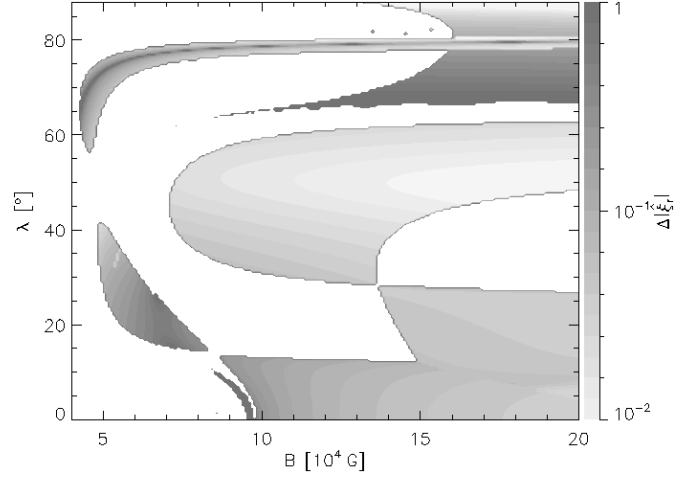}
\caption{
Relative peak-to-peak variation, $\Delta |\hat{\xi}_r|$, of the radial 
envelope. 
For ridge instabilities at low latitudes, $\Delta |\hat{\xi}_r|$ is 
close to unity, and even in the instability island at $\lambda\sim 
15\ldots40\degr$ and low field strengths the resonant wave mode 
coupling leads to considerable values of $\Delta |\hat{\xi}_r|$.
}
\label{ev2d_peak.pic}
\end{figure}

The radial envelope, $|\hat{\xi}_r|$, is used as a measure for the 
probability of loop penetration into the convection zone, subject to 
the normalisation $|\hat{\xi}_r|_\mathrm{max}:= 1$.
The larger the relative peak-to-peak variation,
\begin{equation}
\Delta |\hat{\xi}_r|
:=
\frac{
 |\hat{\xi}_r|_\mathrm{max} - |\hat{\xi}_r|_\mathrm{min}
}{
 |\hat{\xi}_r|_\mathrm{max}
}
\ ,
\label{deltaxir}
\end{equation}
the larger is the possibility for the transition of a loop from the
overshoot region to the convection zone around $\phi_\mathrm{max}$.
Figure \ref{ev2d_peak.pic} shows $\Delta |\hat{\xi}_r|$ as a function 
of $\lambda_0$ and $B_0$.
For $B\gtrsim 10^5\,\mathrm{G}$ the peak-to-peak variation is $\lesssim
5\%$ at intermediate latitudes and up to about $20\%$ near the equator.
Typically $\Delta |\hat{\xi}_r|$ decreases with increasing field 
strengths since the stability properties strongly depend on $\beta 
\delta \propto \delta/B^2$.
High field strengths reduce the effect of the azimuthal 
$\delta$-variation, so that magnetic flux tubes with smaller field 
strengths are, in general, more susceptible to tidal effects. 
Particularly for ridge instabilities and instability islands at low 
field strengths, $\Delta |\hat{\xi}_r|$ is significant and can even 
reach values close to unity, so that radial displacements are almost 
suppressed near the envelope minima.
The approximation used in Eq.~(\ref{absform}) is not applicable here, 
since the dominance of the wave mode $\hat{\xi}_{r,0}$ is not always 
given.
Ridge instabilities are due to resonant interactions of adjacent wave 
modes (see Sec.~\ref{grti}) with comparable amplitudes while the 
amplitudes of all other modes are small.
The case $|\hat{\xi}_{r,0}|, |\hat{\xi}_{r,2}|\gg \hat{\xi}_{r,k}, 
\forall k\neq 0, 2$, for example, yields the radial envelope
\begin{equation}
|\hat{\xi}_r|^2
\propto 
1 
+ 
2
\frac{|\hat{\xi}_{r,0}| |\hat{\xi}_{r,2}|}{
|\hat{\xi}_{r,0}|^2 + |\hat{\xi}_{r,2}|^2}
\cos \left( 2\phi + \alpha_2 - \alpha_0 \right)
\ ,
\label{resoamp}
\end{equation}
with $\alpha_{0/2}= \arg \hat{\xi}_{r,0/2}$.
For $|\hat{\xi}_{r,0}| \approx |\hat{\xi}_{r,2}|$, the radial envelope 
drops to very small values at longitudes which are determined by the 
phase factors $\alpha_{0}$ and $\alpha_{2}$.
The low magnetic field strengths corresponding to these particular 
instabilities imply, however, long growth times which are probably not 
relevant for the evolution of rising flux loops in stars.

%
% Parameterabhaengigkeit
%

\subsection{Parameter dependence}
\label{paab}
We examine the dependence of the stability properties on tidal effects
in various binary systems which are characterised by the orbital 
period, $T$, the mass ratio, $q$, and the mass of the primary star, 
$M_\star$.
Equations (\ref{gtide}) and (\ref{fapprox}) show that the magnitude of
the azimuthal variation caused by tidal effects is governed by the term
\begin{eqnarray}
\epsilon^3 q
& = &
\frac{4\pi^2}{G} 
\frac{q}{1+q}
\frac{r^3}{M_\star T^2}
\nonumber \\
& \simeq &
10^{-2}
\left( \frac{q}{1+q} \right)
\left( \frac{r}{R_{\sun}} \right)^3
\left( \frac{M_\star}{M_{\sun}} \right)^{-1}
\left( \frac{T}{\mathrm{d}} \right)^{-2}
\ .
\label{eps3q}
\end{eqnarray}
Here, the solar-type model is retained, which fixes $M_\star$ and 
$r_0$.
The variation of Eq.~(\ref{eps3q}) due to a different value of $r_0$ 
is negligible since the equilibrium radius is restricted to the rather
thin overshoot layer, which is prescribed by the stellar model.

\begin{figure}
\includegraphics[bb= 35 -1 485 50,width=\hsize]{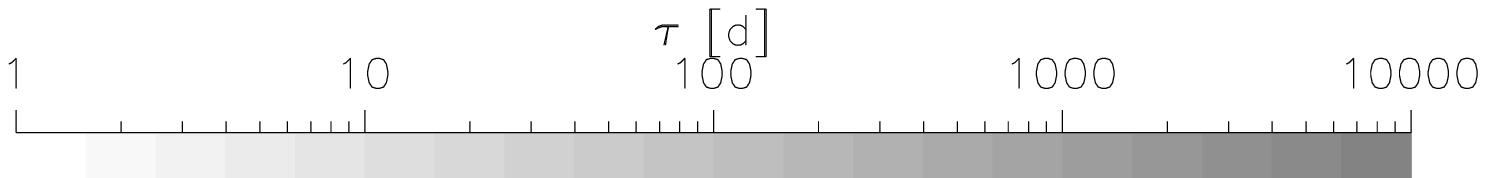} \\
\includegraphics[width=\hsize]{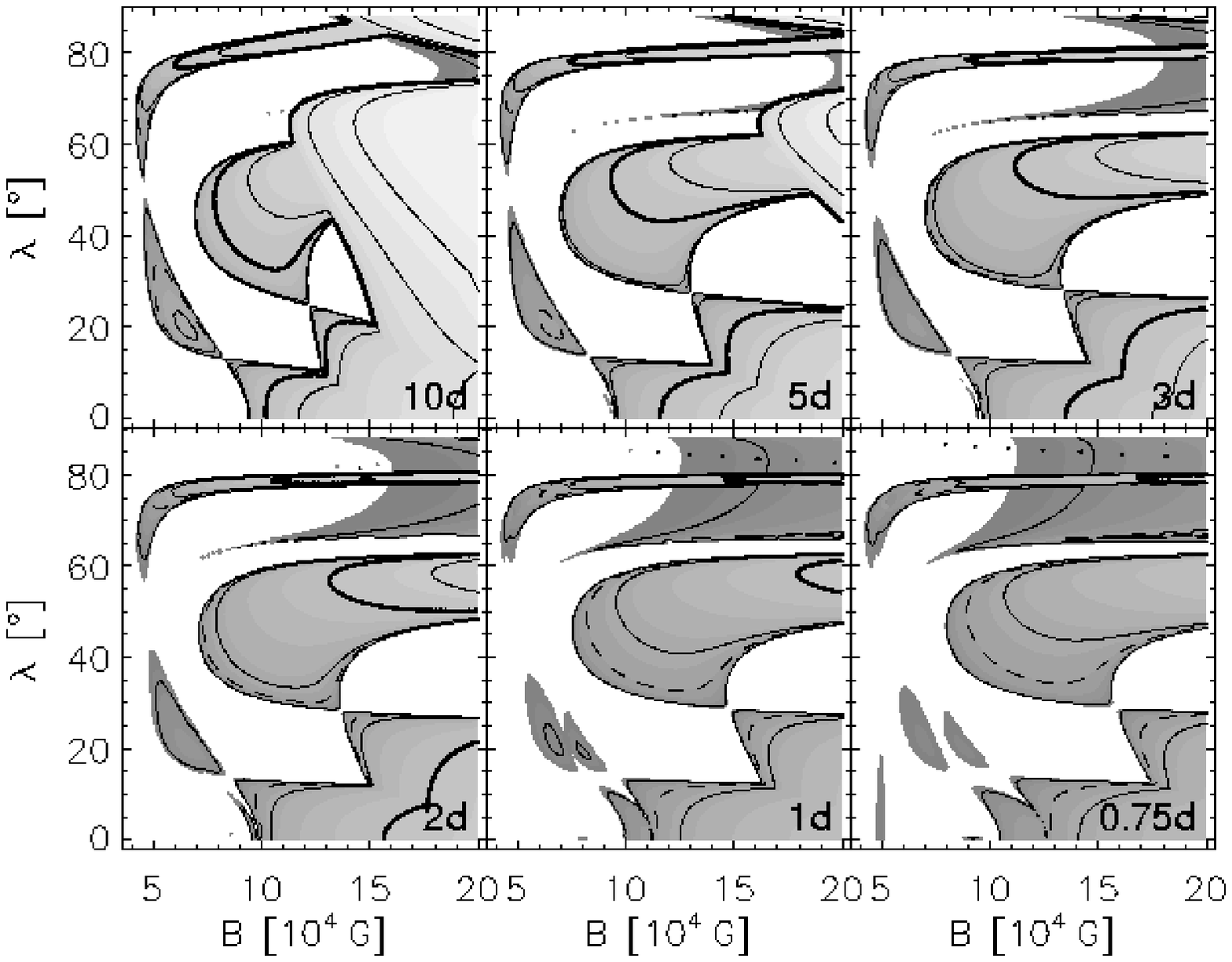}
\caption{
Comparison of growth times, $\tau$, for systems with various orbital 
periods, $T$.
The isolines mark the growth times $\tau= 1,5,10,50,100\ \textrm{(thick
line)}, 500, 1000$ (dashed line), and $5000\,\mathrm{d}$.
Instabilities with $\tau\ge 10^4\,\mathrm{d}$ are omitted.
}
\label{stab_f.pic}
\end{figure}

\begin{figure}
\includegraphics[bb=35 -1 485 54, width=\hsize]{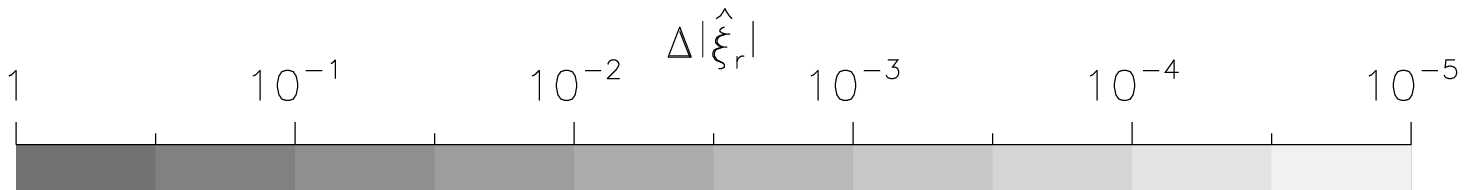} \\
\includegraphics[width=\hsize]{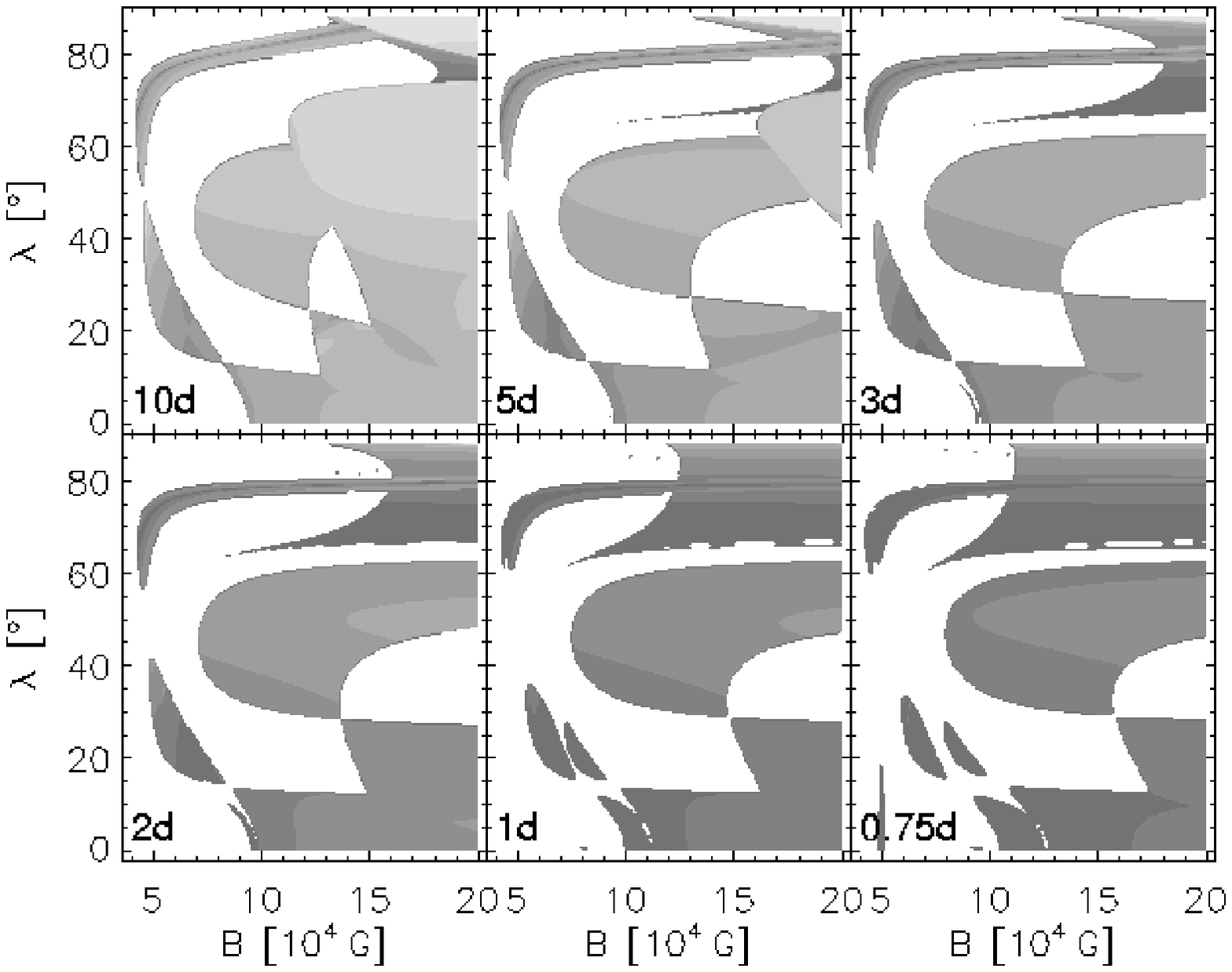}
\caption{
Relative peak-to-peak variation, $\Delta |\hat{\xi}_r|$, of the radial 
envelope for different system periods, $T$.
}
\label{stab_a.pic}
\end{figure}

First we consider binary systems with orbital periods in the range $T= 
0.75\ldots10\,\mathrm{d}$, mass ratio $q= 1$, and, following 
Eq.~(\ref{kepler}), separations $a\simeq 4.4\ldots25\,R_{\sun}$.
Since the expansion parameter covers the range $\epsilon^3= 
4.5\cdot10^{-3}\ldots2.5\cdot10^{-5}$, the analytical approximation for 
the stellar stratification, Eq.~(\ref{fapprox}), is well applicable 
even for the shortest period. 
% Anwachszeiten
A comparison of growth times for different $T$ is shown in 
Fig.~\ref{stab_f.pic}.
A smaller orbital period has a stabilising effect, which is indicated 
by the shifting of corresponding isolines of growth times to higher 
field strengths.
This effect is reinforced by the rotational flattening of the star 
since flux rings at a given, constant equilibrium depth, $r_0$, are
embedded in slightly deeper and more subadiabatic layers of the 
overshoot region.
The tidal effects on $\tau$ are nevertheless small and primarily 
affect the background instabilities, leading to shifts of ridges and 
plateaus.
With decreasing system period, the high-latitude plateau grows toward 
smaller field strengths since the azimuthal variation of the density 
and tidal force along the flux ring become larger; however, the growth 
times remain very large. 
% Phasenlage
The orientations of the radial envelope maxima do not show a 
significant dependence on orbital period and are well represented by 
the reference case with $T= 2\,\mathrm{d}$ (Fig.~\ref{ev2d_phase.pic}).
% Minimalwerte
The relative peak-to-peak variations, $\Delta |\hat{\xi}_r|$, are shown
in Fig.~\ref{stab_a.pic}.
$\Delta |\hat{\xi}_r|$ increases significantly with decreasing orbital 
period, which results in strongly preferred longitudes of loop 
penetrations to the convection zone in short-period systems.
The tidal effects are very effective for flux rings with small field 
strengths, particularly for those which are liable to ridge 
instabilities with $\Delta |\hat{\xi}_r|$ close to unity.
For Parker-type instabilities with short growth times, the relative
peak-to-peak variation remains much smaller.
Figure \ref{stab_a.pic} also shows that the peak-to-peak variation 
strongly decreases with increasing system period for $T\gtrsim 
3\,\mathrm{d}$ and eventually becomes insignificant.

% Massenverhaeltnis
Considering a system with period $T= 2\,\mathrm{d}$, we find that a
variation of the mass ratio in the range $q= 0.1\ldots10$, has only
marginal influence on the stability properties.
In the domain of Parker-type instabilities, the relative changes of the
growth time and of the peak-to-peak variation with respect to the 
reference case are well below $0.1\%$ and $10\%$, respectively.
Furthermore, the orientations of the envelope maxima exhibit hardly any
dependence on the mass ratio.

In summary, the growth times show only a weak dependence on the binary 
parameters, except for the instability background and the rotational 
stabilisation of flux tubes at short system periods. 
The latter effect is mainly due to the conservation of angular momentum
and not innate to the tidal effects.
The relative peak-to-peak variation of the radial envelope exhibits a 
considerable dependence on the orbital period $T$, while the dependence
on the mass ratio $q$ is almost negligible.
Finally, the orientations of the envelope maxima are largely unaffected
by variations of the binary parameters.

%
% Diskussion
%

\section{Discussion}
\label{disc}
Tidal effects and high rotation rates in close binaries cause
characteristic changes of both the equilibrium and the stability
properties of magnetic flux tubes stored in the convective overshoot
region.
Owing to tidal effects, a flux ring in stationary equilibrium
experiences periodic variations of its environment in azimuthal
direction, which modify its characteristic eigenmodes.
To first order, the azimuthal variation of a given quantity depends on 
its local scale height.
Since the stability properties are dominated by the superadiabaticity, 
which is the most strongly depth-dependent quantity in the overshoot 
region, tidal effects are mediated mainly by the azimuthal variation of 
$\delta$.

The growth times of Parker-type instabilities with $\tau\lesssim
1000\,\mathrm{d}$ are only marginally affected by the presence of the
companion star, whereas for instabilities with $\tau\gtrsim
1000\,\mathrm{d}$ the influence of tidal effects is considerably
stronger.
This applies particularly to the instability background, since ridge 
and plateau instabilities owe their existence to the binary character 
and are most susceptible to changes of the tidal effects.
However, since the eigenfrequency is a global parameter describing the
whole flux tube and thus representing an average over the properties of
all tube segments, the dependence of $\omega$ on the binary parameters
is dominated by the axially symmetric influence of stellar rotation and
not so much by the non-axially symmetric tidal interactions: for
axially symmetric changes of the equilibrium configuration, e.g., those
due to the rotational flattening, the changes are identical for each
tube segment and thus may add up to a considerable deviation of
$\omega$, whereas for non-axial, periodic changes along the flux ring,
like those due to tidal effects, the changes of individual tube
segments mutually cancel and eventually average to only minor
deviations of $\omega$.
The growth times of background instabilities are typically several
thousand days and more and therefore very long compared to the system 
period as well as to the growth times which are relevant for rising 
flux tubes in the Sun.
Whether these instabilities are important for loop formation at
preferred longitudes seems questionable, although they possess the most
conspicuously asymmetric eigenfunctions.

\begin{figure}
\begin{center}
\begin{minipage}[][][c]{.315\hsize}
\includegraphics[width=\hsize]{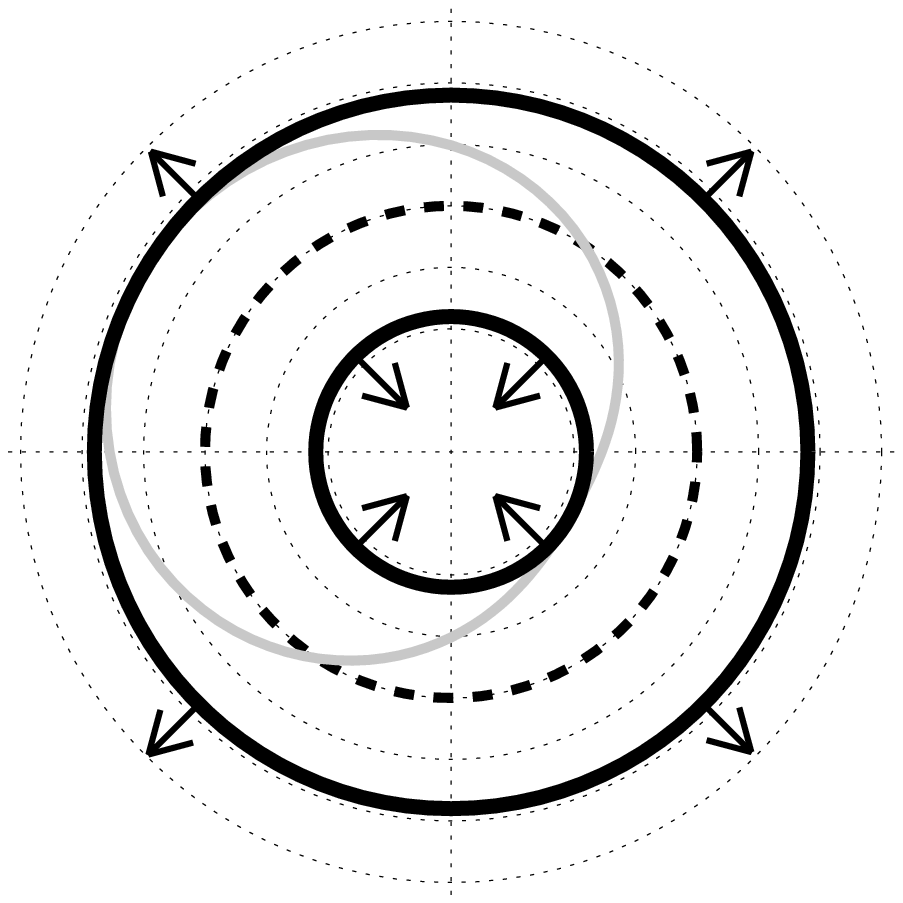}
\end{minipage}
\hfill
\begin{minipage}[][][c]{.315\hsize}
\includegraphics[width=\hsize]{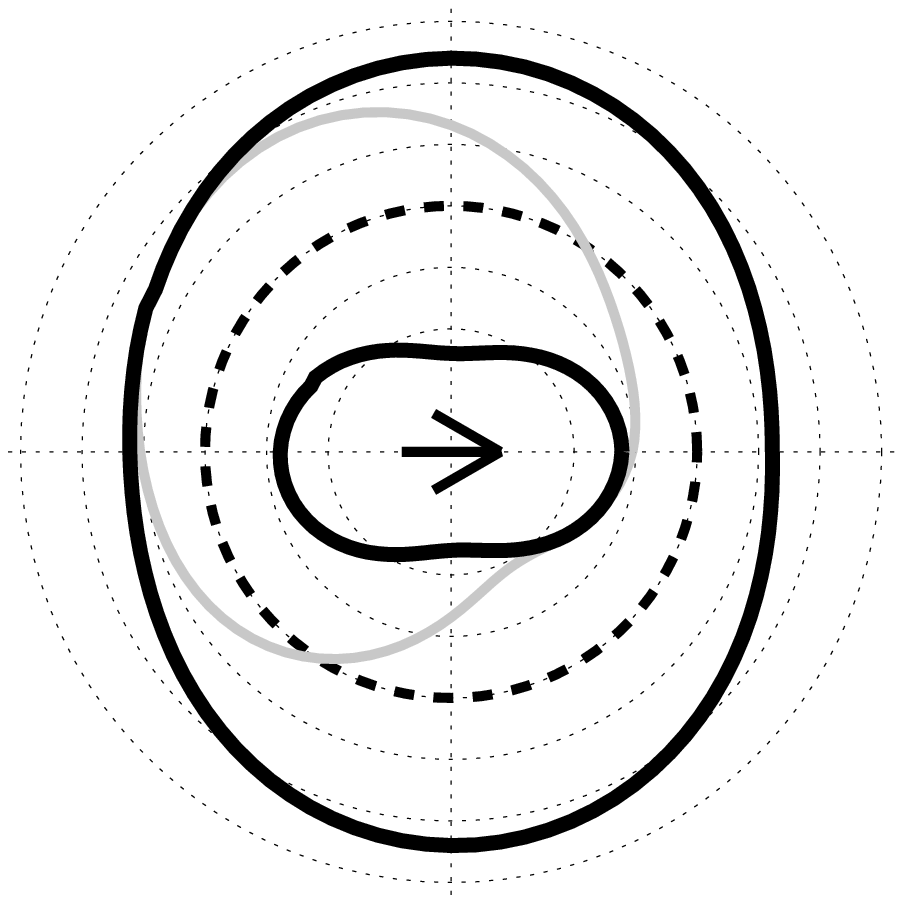}
\newline
\includegraphics[width=\hsize]{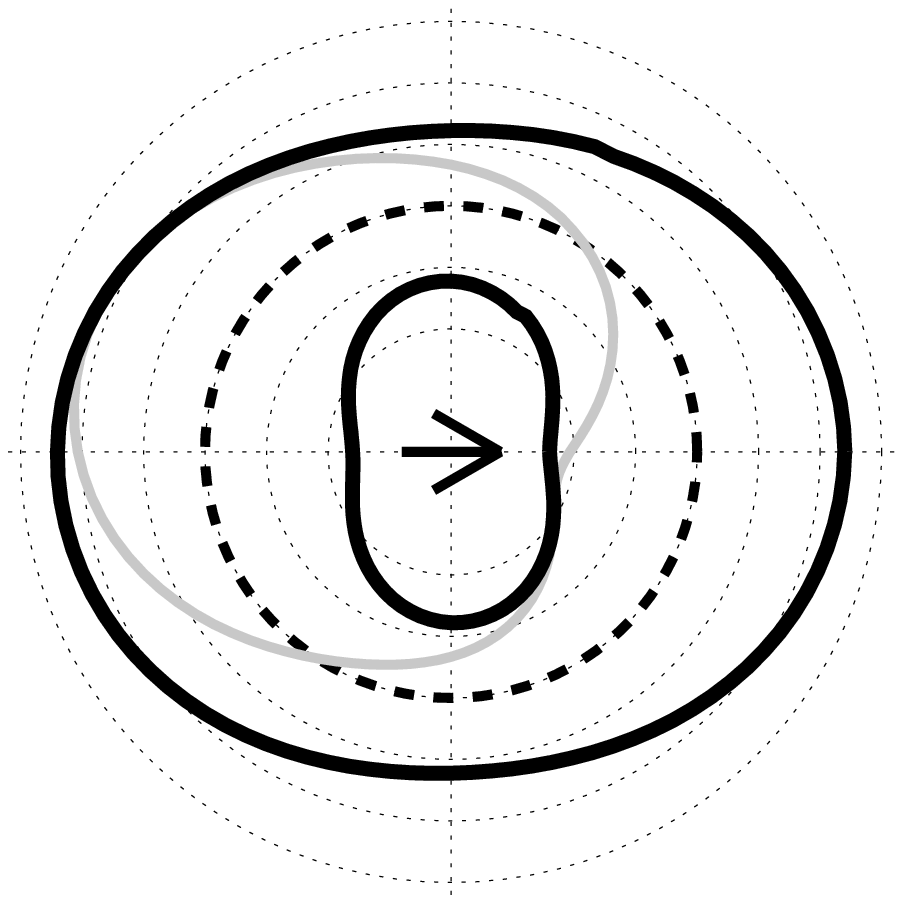}
\end{minipage}
\begin{minipage}[][][c]{.315\hsize}
\includegraphics[width=\hsize]{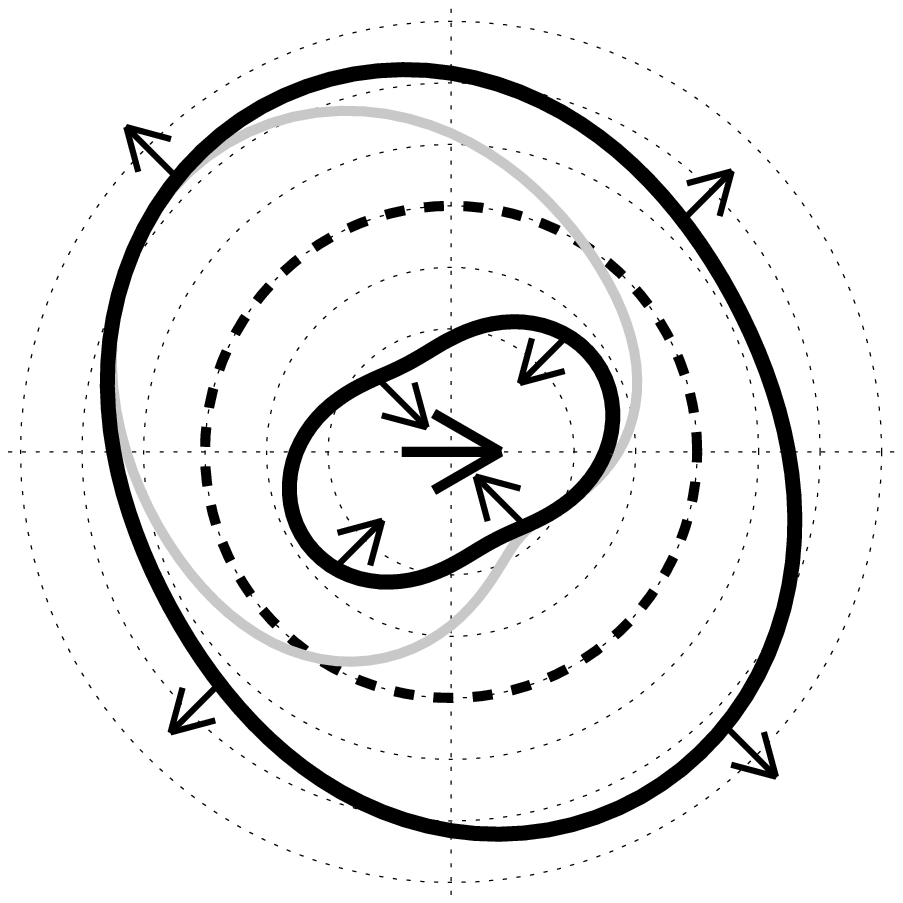}
\end{minipage}
\end{center}
\caption{
Schematic illustration of flux tubes in a single star (\emph{left}) and
in a binary star (\emph{middle} and \emph{right}).
For the latter, the direction to the companion star is indicated by the
central arrow.
The dashed line indicates the equilibrium flux ring, the grey line an 
actual displacement of the perturbed flux tube (here for a wave mode 
with $m= 1$), and the solid black lines the envelope, $|\hat{\xi}_r|$, 
of the eigenfunction.
In the middle panels, two stable modes are shown with the maxima of
their radial envelope located either at the quadrature points
(\emph{upper panel}) or at the line of centres (\emph{lower panel});
unstable modes in binaries typically exhibit phase offsets with respect
to these points (\emph{right}).
}
\label{resuprzp.pic}
\end{figure}

In contrast to the modest effect on the eigenfrequencies and growth 
times, tidal effects can significantly alter the azimuthal structure of
the eigenfunctions since the periodic variations along the tube affect 
the displacement of each segment individually.
The resulting eigenfunction, $\vec{\xi} (\phi)$, characterised by its 
radial envelope $|\hat{\xi}_r|$, consists of a superposition of coupled
wave modes and inherits the $\pi$-periodicity of the azimuthal 
variation of the equilibrium properties.
In contrast to stable eigenmodes, whose envelope maxima are located 
either at the line of centres or at the quadrature points, the maxima 
of unstable eigenmodes typically exhibit a phase offset with respect to
these points; this result is schematically illustrated in 
Fig.~\ref{resuprzp.pic}.
The orientations of the maxima are not unique but depend on the 
equilibrium configuration, predominantly on its latitude and on the 
wave number of the dominating wave mode.
Considering Fig.~\ref{defo.pic}, one might expect that the quadrature 
points would be favoured for rapid loop growth, since they are located 
in less stable layers of the overshoot region and nearest to the 
superadiabatic convection zone.
However, the Parker-type instability is driven by a downflow from the 
loop summit, which increases the density contrast with respect to the 
environment.
Following the equation of continuity, a net downflow implies an 
asymmetry of mass flux at the loop summit, i.e., the amount of plasma 
flowing down in the preceding slope of the loop outbalances the mass 
flux coming up in the following loop slope.
The azimuthal variation of the internal equilibrium density and
velocity, Eqs.~(\ref{rhovar}) and (\ref{velvar}), enhance the asymmetry
in both slopes and thereby favour maximal displacements at intermediate
longitudes.
In contrast, the extrema of stable eigenmodes prefer either the
quadrature points or the directions along the line connecting both
stellar centres since at these 'symmetry points' (concerning the
internal equilibrium density and velocity structure) the downflow and
thus buoyancy effects can be minimised.

The peak-to-peak variation, $\Delta |\vec{\hat{\xi}}_r|$, serves as a 
measure for the probability of loop penetration into the convection 
zone, based on the assumption that the loop segments lifted furthest 
toward the superadiabatic region are the most likely to leave the 
overshoot region.
The radial envelope determines the displacement of the perturbation
wave and thus the formation of rapidly growing loops inside the
convection zone is favoured around its maxima at $\phi_\mathrm{max}$
and $\phi_\mathrm{max} + \pi$.
In the case of close binaries with orbital periods of a few days, the
peak-to-peak variation of fast growing Parker-type instabilities is of 
the order of several percent, up to about $20\%$.
In the case of ridge instabilities, radial displacements are strongly
suppressed at some longitudes, giving rise to distinct preferred
longitudes in the vicinity of $\phi_\mathrm{max}$ and
$\phi_\mathrm{max} + \pi$.
Although it can not be excluded that ridge instabilities contribute to
the non-uniform spot distribution, their importance for the formation 
of preferred longitudes at the stellar surface is questionable since 
the relevant equilibrium configurations are restricted to rather small 
parameter domains $(B_0,\lambda_0)$, where instabilities have large 
growth times.
Based on the parameter study in Sect.~\ref{paab}, we would predict that
preferred longitudes appear (in binaries with solar-type components) 
for orbital periods $T\lesssim 5\,\mathrm{d}$, which is in agreement 
with the results of \citet{1995AJ....109.2169H}.
However, in some binaries with longer periods (consisting of giant
components) preferred longitudes have also been observed 
\citep{1998A&A...336L..25B}.

The linear stability analysis establishes the existence of preferred 
longitudes at the bottom of the convection zone.
However, the determination of the actual longitude where an 
\emph{individual} loop enters the convection zone is not possible since
our procedure only allows \emph{statistical} predictions.
The tendency to constitute preferred longitudes is instead discernible 
for a large ensemble of similar flux tubes, supposed that the 
environment conditions in the overshoot region, e.g., the amount and 
distribution of overshooting convective motions, are uniform and 
unaltered for a sufficiently long time.
Preferred longitudes at the bottom of the convection zone are not
necessarily simply related to the observed preferred longitudes of
starspots at stellar surfaces, because the evolution of rising loops
through the convection zone is still subject to tidal
effects\footnote{If the tube evolution would be linear throughout the
convection zone, the maxima of the radial envelope would, in fact, also
determine preferred longitudes of flux eruption at the surface.}.
The preferred longitudes of flux eruption at the stellar surface based
on the flux tube model described above have to be determined by 
non-linear numerical calculations.
Results of such calculations are presented in the subsequent paper II.

%
% Schlussfolgerungen
%

\section{Conclusion}
\label{conc}
The equilibrium and stability properties of toroidal magnetic flux
tubes in the overshoot region of an active component of a close binary
system are considerably affected by tidal effects.
The breaking of the axial symmetry due to the presence of the companion
star leads to azimuthal variations along the tube of the equilibrium 
configuration and of the probability of the penetration of unstable
loops to the convection zone.
This demonstrates the existence of preferred longitudes \emph{at the 
bottom of the convection zone}, which are well pronounced for orbital 
periods of a few days but strongly fade with increasing period.
The orientations of the preferred longitudes depend on the equilibrium 
latitude, the field strength, and the dominating wave mode of the 
unstable eigenmode.
Despite their rather small magnitude, tidal effects are nevertheless
capable of influencing the dynamics of magnetic flux tubes considerably
because they are enhanced by the sensitive dependence of the stability
properties on the superadiabaticity of the local environment and by the
resonant coupling between interacting wave modes.

\begin{acknowledgements}
Volkmar Holzwarth thanks Prof.~S.~K.~Solanki and
Prof.~F.~Moreno-Insertis for valuable discussions and the
Max-Planck-Institut f\"ur Aeronomie in Katlenburg-Lindau and the
Kiepenheuer--Institut f\"ur Sonnenphysik in Freiburg/Brsg.~for
financial support during the accomplishment of this work.
\end{acknowledgements}

\appendix

\section{Numerical treatment} 
\label{nume}
The perturbations of all physical and geometrical quantities (magnetic 
field, flow velocity, density, position, curvature,\ldots) arising from
the small perturbations of an equilibrium flux ring are expressed in 
terms of the displacement vector $\vec{\xi} (\phi)$ 
\citep[see, e.g.,][]{1993GAFD...72..209}.
The substitution of the corresponding expressions in the equation of 
motion followed by a linearision with respect to $\vec{\xi}$ yields 
Eq.~(\ref{hlinsys}), with the coefficient matrices\footnote{In the case
$\beta\gg 1$.}
\begin{eqnarray}
\mathcal{M}_{\phi t}
& = &
-
\frac{1}{\Omega}
\mathcal{E}
\label{matphit}
\\
\mathcal{M}_{tt}
& = &
-
\frac{1}{4}
\frac{1}{\Omega^2}
\frac{1 - M_\alpha^2}{M_\alpha^2}
\mathcal{E}
\label{mattt}
\\
\mathcal{M}_\phi
& = &
\left(
 \mathcal{F}_1
 \frac{g_\mathrm{n}}{g_*}
 -
 1
\right)
\mathcal{Y}
+
\mathcal{F}_1
\frac{g_\mathrm{b}}{g_*}
\mathcal{Z}
\label{matphi}
\\
\mathcal{M}_t
& = &
\frac{1}{2}
\frac{1}{\Omega}
\frac{1 + M_\alpha^2}{M_\alpha^2}
\mathcal{Y}
\label{matt}
\\
\mathcal{M}_\xi
& = &
\mathcal{F}_1
\frac{1}{g_*}
\left(
 \begin{array}{ccc}
  g_\mathrm{t}'
 &
  -g_\mathrm{t}
 &
  0
 \\
  g_\mathrm{t}
 &
  2
  g_\mathrm{n}
 &
  g_\mathrm{b}
 \\
  0
 &
  g_\mathrm{b}
 &
  0
 \end{array}
\right)
+
\mathcal{F}_3
\frac{1}{g_*^2}
\left(
 \begin{array}{ccc}
  g_\mathrm{t}
  g_\mathrm{t}
 &
  g_\mathrm{t}
  g_\mathrm{n}
 &
  g_\mathrm{t}
  g_\mathrm{b}
 \\
  g_\mathrm{n}
  g_\mathrm{t}
 &
  g_\mathrm{n}
  g_\mathrm{n}
 &
  g_\mathrm{n}
  g_\mathrm{b}
 \\
  g_\mathrm{b}
  g_\mathrm{t}
 &
  g_\mathrm{b}
  g_\mathrm{n}
 &
  g_\mathrm{b}
  g_\mathrm{b}
 \end{array}
\right)
\nonumber \\
& &
{}
+
\mathcal{F}_2
\frac{1}{g_*}
\left(
 \begin{array}{ccc}
  g_\mathrm{t}
 &
  g_\mathrm{n}
 &
  g_\mathrm{b}
 \\
  0
 &
  0
 &
  0
 \\
  0
 &
  0
 &
  0
 \end{array}
\right)
\ ,
\label{matxi}
\end{eqnarray}
where the constant matrices 
\begin{equation}
\mathcal{E}
=
\left(
 \begin{array}{ccc}
  1 & 0 & 0 \\ 0 & 1 & 0 \\ 0 & 0 & 1
 \end{array}
\right)
\ , \quad
\mathcal{Y}
=
\left(
 \begin{array}{ccc}
  0 & 1 & 0 \\ -1 & 0 & 0 \\ 0 & 0 & 0
 \end{array}
\right)
\ , \quad
\mathcal{Z}
=
\left(
 \begin{array}{ccc}
  0 & 0 & 1 \\ 0 & 0 & 0 \\ -1 & 0 & 0
 \end{array}
\right)
\ ,
\end{equation}
and the abbreviations
\begin{eqnarray}
\mathcal{F}_1
& = &
\frac{1}{\left(1-M_\alpha^2\right)}
\frac{\rho_\mathrm{e}}{p_\mathrm{e}}
\frac{1}{\gamma} 
\frac{g_*}{\kappa}
\\
\mathcal{F}_2
& = &
-
\mathcal{F}_1^2
\left[
  \left(
    1
    -
    \nabla
  \right)
  M_\alpha^2
  +
  \left(
    1
    -
    M_\alpha^2
  \right)
  \nabla
\right]
\gamma
\frac{g_\mathrm{t}}{g_*}
\\
\mathcal{F}_3
& = &
\mathcal{F}_1^2
\frac{\gamma^2}{2}
\Delta
\left(1-M_\alpha^2\right)
\\
\Delta 
& = &
\beta 
\delta 
- 
\frac{2}{\gamma} 
\left( 
  \frac{1}{\gamma} 
  - 
  \frac{1}{2} 
\right)
\end{eqnarray}
have been used.
The components $g_\mathrm{x}= \left( \vec{g}_\mathrm{eff} \cdot \vec{x}
\right), \vec{x}\in \left[ \vec{t}, \vec{n}, \vec{b} \right]$ of the 
effective gravitational acceleration $\vec{g}_\mathrm{eff}$ are 
expressed in the co-moving trihedron.

Owing to the azimuthal variation of the equilibrium quantities induced
by the presence of the companion star, the coefficient matrices
(\ref{mattt}) -- (\ref{matxi}) can be expressed in the form of
Eq.~(\ref{matvar}) and comprise small contributions which are, in
lowest order, $\pi$-periodic in the longitude $\phi$.
Following the ansatz in Eq.~(\ref{ampans}), these contributions cause
the coupling of wave modes $\hat{\xi}_\mathrm{n}$ with wave numbers $n=
m, m\pm2, m\pm4,\ldots$, which is expressed either by the 3-term
recursion formula in Eq.~(\ref{recu}) or by the Hill-type algebraic
system
\begin{equation}
\underbrace{
\left(
 \begin{array}{*{5}{c}}
   \mathcal{L}_{m-2K} & \mathcal{C}_{m-2K} & \mathcal{R}_{m-2K}
  \\
   & \hspace{-1.5cm} \ddots & \hspace{-1.5cm} \ddots & 
     \hspace{-1.5cm} \ddots & \vec{O}
  \\
   & \mathcal{L}_m & \mathcal{C}_m & \mathcal{R}_m 
  \\
   \vec{O} & & \hspace{-1.5cm} \ddots & \hspace{-1.5cm} \ddots & 
    \hspace{-1.5cm} \ddots
  \\
   & & \mathcal{L}_{m+2K} & \mathcal{C}_{m+2K} & \mathcal{R}_{m+2K} 
 \end{array}
\right)
}_{\mathcal{H} (\omega)}
\left(
 \begin{array}{c}
  \vec{\hat{\xi}}_{-2K} \\ \vdots \\ \vec{\hat{\xi}}_{0} \\ 
  \vdots \\ \vec{\hat{\xi}}_{2K}
 \end{array}
\right)
=
0
\ ,
\label{nuve:hillcut}
\end{equation}
with $K\rightarrow\infty$ and the block matrices
\begin{eqnarray}
\mathcal{L}_n
& = &
\frac{1}{2}
\left[
 \left( n - 2 \right) 
 \left( 
  i \mathcal{M}_{\phi,c} + \mathcal{M}_{\phi,s} 
  -
  \omega 
  \left( \mathcal{M}_{\phi t,c} - i \mathcal{M}_{\phi t,s} \right) 
 \right)
\right.
\nonumber \\ & &
\left.
 + 
 \mathcal{M}_{\xi,c} 
 - 
 i \mathcal{M}_{\xi,s} 
 + 
 \omega 
 \left( i \mathcal{M}_{t,c} + \mathcal{M}_{t,s} \right)
\right.
\nonumber \\ & &
\left.
 - 
 \omega^2 
 \left( \mathcal{M}_{tt,c} - i \mathcal{M}_{tt,s} \right)
\right]
\ ,
\\
\mathcal{R}_n
& = &
\frac{1}{2}
\left[
 \left( n + 2 \right)
 \left(
  i \mathcal{M}_{\phi,c} - \mathcal{M}_{\phi,s}
  -
  \omega
  \left( \mathcal{M}_{\phi t, c} + i  \mathcal{M}_{\phi t,s} \right)
 \right)
\right.
\nonumber \\ & &
\left.
 +
 \mathcal{M}_{\xi,c} 
 + 
 i \mathcal{M}_{\xi,s} 
 + 
 \omega
 \left( i \mathcal{M}_{t,c} - \mathcal{M}_{t,s} \right)
\right.
\nonumber \\ & &
\left.
 - 
 \omega^2
 \left( \mathcal{M}_{tt,c} + i \mathcal{M}_{tt,s} \right)
\right]
\ ,
\end{eqnarray}
and
\begin{eqnarray}
\mathcal{C}_n
& = &
- n^2 \mathcal{E} 
- n \omega \mathcal{M}_{\phi t,0}
- \omega^2 \mathcal{M}_{tt,0}
+ \mathcal{M}_{\xi,0}
\nonumber \\ & &
+ i \left( n \mathcal{M}_{\phi,0} + \omega  \mathcal{M}_{t,0} \right)
\ .
\end{eqnarray}
The eigenfrequencies $\omega$ are determined by the roots of the
dispersion relation, $\det \left[ \mathcal{H} (\omega) \right]= 0$.
A treatment of the case $K\rightarrow \infty$ is not practical, but 
since the coupling between adjacent wave modes is small [$\mathcal{L},
\mathcal{R}\sim \mathcal{O}(\epsilon^3)$], the wave mode spectrum can 
be truncated at a finite number of constituents, here $K= 3$. 
A verification of the eigenfrequency and the determination of the 
eigenfunctions $\vec{\xi} (\phi)$ is based on the following approach.
Let the independent solutions $x_{(i)} (\phi), i=1,\ldots,n$ of a 
system 
\begin{equation}
x' = \mathcal{A} (\phi) x
\qquad \textrm{with} \quad
\mathcal{A} (\phi + \pi)= \mathcal{A} (\phi)
\label{glsy}
\end{equation}
represent the fundamental matrix $\mathcal{X} (\phi)= \left( x_{(1)},
\ldots, x_{(n)} \right)$ with $\det[ \mathcal{X} (\phi) ] \neq 0$.
Furthermore, we have $\mathcal{X} (\phi + 2 \pi)= \mathcal{X} (\phi) 
\mathcal{M}$, where $\mathcal{M}$ is the constant \emph{transition 
matrix} with $\det(\mathcal{M})\neq 0$.
Since an arbitrary solution of Eq.~(\ref{glsy}) can be represented by a
constant vector $\vec{c}$ in the form $x (\phi)= \mathcal{X} (\phi) 
\vec{c}$, it follows that $x (\phi + 2 \pi)= \mathcal{X} (\phi + 2 \pi) 
\vec{c}= \mathcal{X} (\phi) \mathcal{M} \vec{c}$.
Periodic solutions with $x (\phi + 2 \pi) \equiv x (\phi)$ thus require 
$\mathcal{M} \vec{c}= \vec{c}$ and an eigenvalue of the transition 
matrix $\mathcal{M}$ equal to unity.
If the fundamental matrix fulfils the initial condition $\mathcal{X} 
(0)= \mathcal{E}$, the transition matrix is calculated by integrating
over a period of $2\pi$, i.e., $\mathcal{X} (2 \pi)= \mathcal{M}$.
In our case, the transition matrix $\mathcal{M}$ and its eigenvalues 
depend on the eigenfrequency, $\omega$.
If an eigenfrequency determined by the dispersion relation above does 
not entail $\mathcal{M} \vec{c}= \vec{c}$, $\omega$ is iteratively
improved and the eigenfunction $\vec{\xi} (\phi)$ determined.

\end{document}